\documentclass[pra,amsmath,amssymb,aps,graphicx,onecolumn,reprint]{revtex4-2}

\usepackage{graphicx}
\usepackage{physics}
\usepackage{color}
\usepackage{mathrsfs}

\begin{document}
\newcommand{\dif}{\mathrm{d}}
\makeatletter
\renewcommand\@biblabel[1]{#1.}
\makeatother

 \title{A flying Schrödinger cat in multipartite entangled states}

\author{Zhiling Wang}
\altaffiliation{These two authors contributed equally to this work.}

\author{Zenghui Bao}
\altaffiliation{These two authors contributed equally to this work.}

\author{Yukai Wu}

\author{Yan Li}

\author{Weizhou Cai}

\author{Weiting Wang}

\author{Yuwei Ma}

\author{Tianqi Cai}

\author{Xiyue Han}

\author{Jiahui Wang}

\author{Yipu Song}

\author{Luyan Sun}

\author{Hongyi Zhang}
\email{hyzhang2016@tsinghua.edu.cn}

\author{Luming Duan}
\email{lmduan@tsinghua.edu.cn}

\affiliation{Center for Quantum Information, Institute for Interdisciplinary Information Sciences, Tsinghua University, Beijing 100084, PR China}

\date{\today}

\begin{abstract}
Schrödinger's cat originates from the famous thought experiment querying the counterintuitive quantum superposition of macroscopic objects. As a natural extension, several `‘cats" (quasi-classical objects) can be prepared into coherent quantum superposition states, which is known as multipartite cat states demonstrating quantum entanglement among macroscopically distinct objects. Here we present a highly scalable approach to deterministically create flying multipartite Schrödinger cat states, by reflecting coherent state photons from a microwave cavity containing a superconducting qubit. We perform full quantum state tomography on the cat states with up to four photonic modes and confirm the existence of quantum entanglement among them. We also witness the hybrid entanglement between discrete-variable states (the qubit) and continuous-variable states (the flying multipartite cat) through a joint quantum state tomography. Our work provides an enabling step for implementing a series of quantum metrology and quantum information processing protocols based on cat states. 
\end{abstract}

\maketitle 
\section{Introduction}

In Schrödinger’s thought experiment, a cat would be in a peculiar mixture of being dead and alive if it is entangled with an atom~\cite{Sch1935}. In quantum experiments, the cat is usually emulated by quantum superposition of macroscopically distinct states, for example, superposition of two coherent states with opposite phases~\cite{Stoler1986}. Once being entangled with an atom, such a system equivalently forms a Schrödinger's cat. Preparing Schrödinger cat states has attracted wide research interest from testing quantum foundations to demonstrating the increasing controllability of modern quantum systems. Over the decades, Schrödinger cat states have been successfully prepared in various physical systems, including vibrational states of a trapped ion~\cite{Wineland2013}, propagating photon modes~\cite{Ourjoumtsev2006,Ourjoumtsev2007,Takahashi2008,Namekata2010,Alexander2017,Schoelkopf17,Rempe2019,Wallraff2020,lewenstein2021generation,sychev2017enlargement, morin2014remote, jeong2014generation, neergaard2010optical} and microwave photons confined in superconducting cavities coupled with either Rydberg atoms~\cite{Haroche2013,Haroche2008} or superconducting qubits~\cite{Schoelkopf2013}. By entangling with more continuous-variable modes, a multipartite Schrödinger's cat can be obtained as $\mathscr{N}(\ket{\alpha}^{\otimes n} \pm \ket{-\alpha}^{\otimes n})$, where $\ket{\alpha}$ represents a coherent state, $\mathscr{N}$ is a normalization factor, and $n$ represents the number of photonic modes (the number of entangled parties), essentially describing quantum entanglement of coherent states~\cite{Sanders2012}. It is of great importance for fundamental tests of the quantum non-locality~\cite{Philippe2003}, and for implementation of quantum metrology~\cite{Munro2002}, quantum information processing~\cite{Jeong2001,Ralph2003} and quantum network~\cite{Loock2008,Sangouard2010}. The preparation of bipartite cat states has been demonstrated in circuit quantum electrodynamics systems for stationary microwave photonic modes in resonators~\cite{wang2016}, or through a nondeterministic photon subtraction for propagating optical photons~\cite{Grangier2009}. However, those protocols are hard to be extended for generation of multipartite cat states, limited either by the requirement of intermediate stationary cat states or the nondeterministic nature of the scheme.

In this work, we demonstrate a highly scalable approach to generate multipartite cat states~\cite{Duan04cz,Duan05cat} for itinerant microwave photons, or flying cat states, by sequentially reflecting coherent state microwave pulses from a resonator containing a superconducting qubit. We prepare even and odd cat states containing up to four photonic modes, and further demonstrate the possibility to generate more superposition and entanglement structures by coherent control of the qubit state. The full quantum state tomography is performed on the itinerant photonic modes to verify the existence of quantum entanglement of the multipartite cat states. Hybrid entanglement between the qubit and the flying cat states is also confirmed through a joint quantum state tomography. We note that the fidelity and the scale of the cat states prepared with our method are mainly limited by the resonator internal loss and the qubit decoherence, which can be alleviated to prepare larger scale multipartite cat states. Our work thus presents a highly scalable scheme for the deterministic generation of multipartite cat states and provides an important enabling tool to realize various quantum metrology~\cite{Fink2020} and quantum information processing protocols based on continuous-variable cat states~\cite{Loock2016,Jiang2017prl}.

\begin{figure*}[!tbp]
\centering
\includegraphics[width=0.7\linewidth]{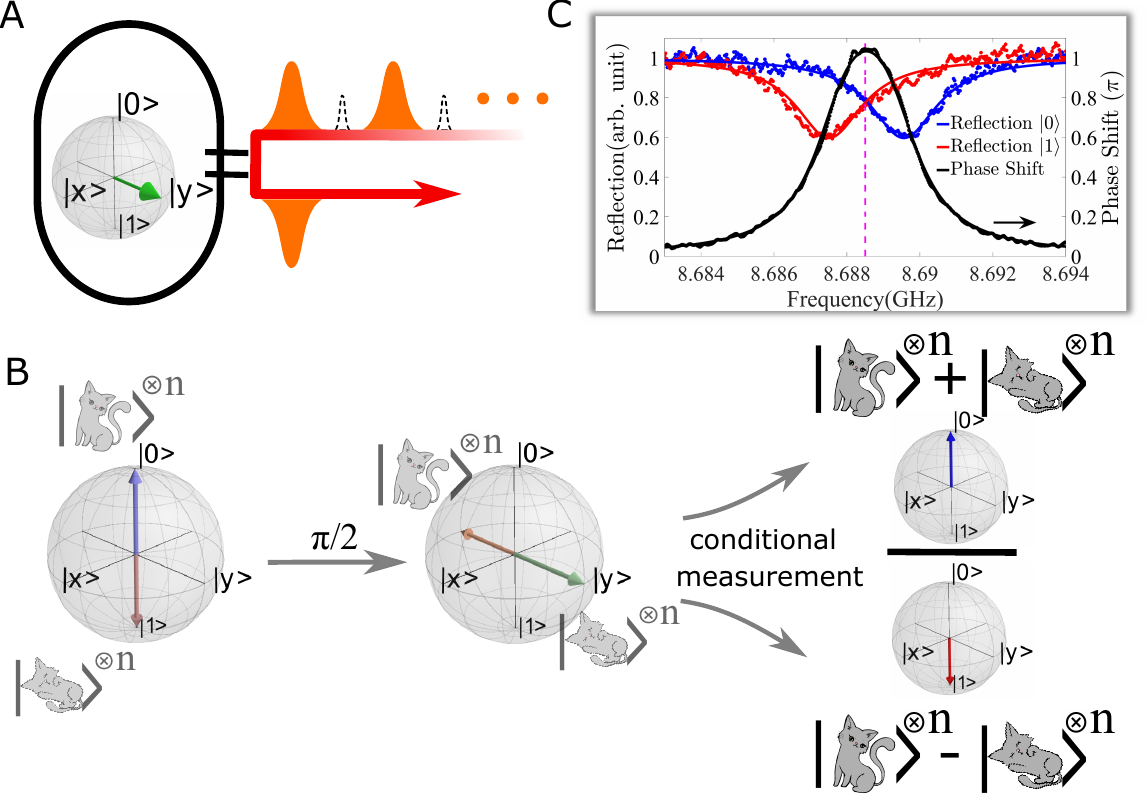}
\caption{\textbf{The protocol for the preparation of flying multipartite cat states.} \textbf{A,} We use a microwave cavity~\cite{Paik2011,Sun2019} dispersively coupled to a superconducting qubit to deterministically generate multipartite cat states of itinerant microwave photons. We first set the qubit to a superposition state $(\ket{0}+\ket{1})/\sqrt{2}$, and send itinerant microwave photons in the coherent state (orange pulses) to the microwave cavity. The reflected photons would acquire a conditional $\pi$ phase shift depending on the state of the qubit, and thus we would have $\ket{\alpha}$ ($\ket{-\alpha}$) if the qubit state is in $\ket{0}$ ($\ket{1}$). By sending a train of coherent state photon pulses which are sequentially reflected from the cavity, we would have a multipartite entangled state $(\ket{0}\ket{\alpha}^{\otimes n}+\ket{1}\ket{-\alpha}^{\otimes n})$. \textbf{B} shows the state evolution during the preparation of the multipartite cat states, where we use a cat to represent the quasi-classical coherent states. Starting with the multipartite entangled state, the qubit is rotated with a $\pi$/2 pulse, resulting in a state as $((\ket{0}+\ket{1})\ket{\alpha}^{\otimes n}+(\ket{0}-\ket{1})\ket{-\alpha}^{\otimes n})$. Then, the qubit state is mapped to either $\ket{0}$ or $\ket{1}$ with a projective measurement. The photon state conditioned on the qubit state of $\ket{0}$ (or $\ket{1}$) is thus an even n-partite cat state $\mathscr{N}(\ket{\alpha}^{\otimes n}+\ket{-\alpha}^{\otimes n})$(or an odd n-partite cat state $\mathscr{N}(\ket{\alpha}^{\otimes n}-\ket{-\alpha}^{\otimes n})$). It is worth mentioning that more superposition and entanglement structures can be generated by varying the amplitudes of input coherent pulses, or by applying specific unitary qubit rotations between the adjacent reflections of the photon pulses, illustrated in \textbf{A} as the pulses with dotted envelope. \textbf{C} shows the measured cavity reflection spectra when the qubit is in either $\ket{0}$ (blue) or $\ket{1}$ (red). The resulting phase difference of the qubit-state-dependent reflections is shown in black, which reaches $\pi$ at the bare cavity frequency $\omega_c$. The dots are measured results and the solid lines are theoretical fitting results. The dashed line indicates the frequency of the input photons.}
\label{illustration}
\end{figure*}

\section{Results}
\subsection{The protocol}

We consider a superconducting microwave cavity (resonator) dispersively coupled to a transmon qubit~\cite{JC1,JC2}. The Hamiltonian of the system can be written as $H/\hbar=(\omega_r+\chi\sigma_z)a^\dagger a+\omega_q\sigma_z/2$, where $\omega_r$ is the cavity frequency, $a$ ($a^\dagger$) is the annihilation (creation) operator of the cavity mode, $\omega_q$ is the qubit frequency, $\sigma_z$ is Pauli operator of the qubit, and $\chi$ represents dispersive shift induced by the interaction between the qubit and the cavity. The cavity resonance frequency depends on the state of the qubit due to the dispersive term. As shown in Fig.~\ref{illustration}C, we tune the cavity linewidth $\kappa$ to approximately $2|\chi|$, resulting in qubit-state-dependent reflections with the same strength but the opposite phases if the input photon frequency is at $\omega_r$~\cite{Nakamura2018}.
Consequently, if a train of $n$ coherent state microwave photon pulses at $\omega_r$ are sent to the cavity, the reflected photon state can be expressed as $\ket{\alpha}^{\otimes n}$ (or $\ket{-\alpha}^{\otimes n}$) for qubit in the ground state $\ket{0}$ (or in the excited state $\ket{1}$).
If the qubit is prepared in a quantum superposition state $(\ket{0}+\ket{1})/\sqrt{2}$, the combined system is then in a multipartite Schrödinger cat state in the form~\cite{Duan05cat}

\begin{equation}
\frac{1}{\sqrt{2}}(\ket{0}\ket{\alpha}^{\otimes n}+\ket{1}\ket{-\alpha}^{\otimes n}).
\label{qph-entan}
\end{equation}
We can further apply another $\pi/2$ rotation along $-$y-axis $R_{-y}(\pi/2)$ after the photon reflections, yielding the state 
\begin{equation}
\begin{split}
&\frac{1}{2}[(\ket{0}-\ket{1})\ket{\alpha}^{\otimes n}+(\ket{0}+\ket{1})\ket{-\alpha}^{\otimes n}]\\
&=\frac{1}{2}[\ket{0}(\ket{\alpha}^{\otimes n}+\ket{-\alpha}^{\otimes n})-\ket{1}(\ket{\alpha}^{\otimes n}-\ket{-\alpha}^{\otimes n})].
\end{split}
\label{q-ph-pi2}
\end{equation}
Therefore, an even or odd multipartite cat state of microwave photons $\mathscr{N}(\ket{\alpha}^{\otimes n}\pm\ket{-\alpha}^{\otimes n})$ can be obtained by projecting the qubit state to either $\ket{0}$ or $\ket{1}$. It should be emphasized that obtaining well defined multipartite cat states requires carefully determined pulse lengths and time intervals between the pulses, to fit the
bandwidth of the cavity and to ensure that the photonic modes can be well distinguished in the time domain, respectively.

To generate more superposition and entanglement structures of the coherent states, one can apply some specific unitary qubit rotations between the adjacent reflections of the photon pulses, as illustrated in Fig.~\ref{illustration}A. The additional qubit rotations effectively prepare different forms of multipartite Schrödinger cat state compared with Eq.~\ref{qph-entan}. Detailed discussions can be found in Section.~\uppercase\expandafter{\romannumeral1} D of the Supplementary Materials. Additionally, since the coherent pulses are generated externally by a microwave signal source, it is convenient to utilize different amplitudes and shapes for the pulses, and thus obtain well tunable sizes for each component of the multipartite cat states.

\begin{figure*}[!tbp]
\centering
\includegraphics[width=1\linewidth]{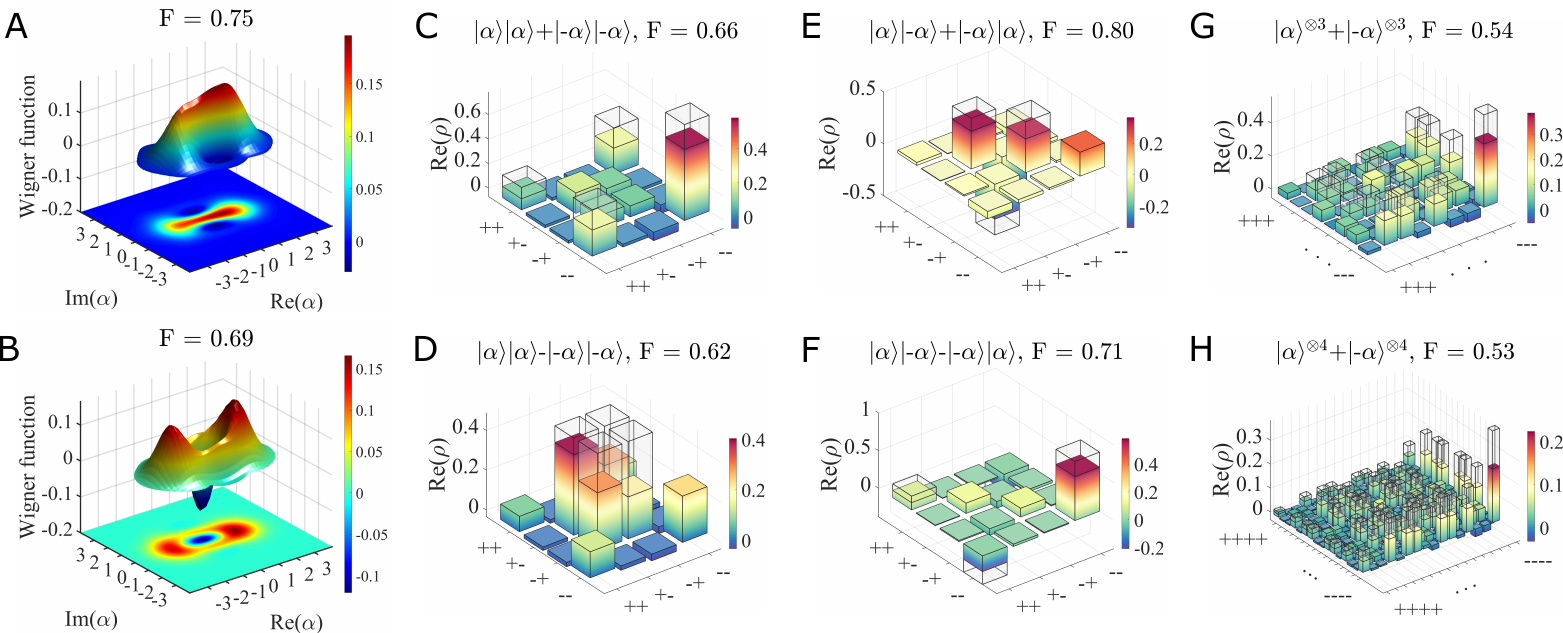}
\caption{\textbf{Multipartite cat states.} \textbf{A, B,} Three-dimensional plots of Wigner functions of the single-party even and odd cat states with $\alpha = 1.07$, obtained when the qubit is measured in $\ket{0}$ and $\ket{1}$, respectively. \textbf{C, D,} The real part of the density matrices for even and odd bipartite cat states with $\alpha = 0.72$ when the qubit is projected in $\ket{0}$ and $\ket{1}$, respectively. The itinerant photon states are reconstructed with a maximum likelihood quantum state estimation in Fock basis. For clarity, we plot the obtained density matrices in the cat basis $|\pm\rangle \propto |\alpha\rangle \pm |-\alpha\rangle$ with small population outside these states not shown. \textbf{E, F,} The real part of the density matrices for bipartite cat states when inserting a qubit $\pi$ rotation between the two successive reflections of microwave photons, conditioned on the qubit measured in $\ket{0}$ and $\ket{1}$, respectively. The $\pi$ rotation refocuses some of the qubit dephasing, resulting in improved fidelities of the generated cat states. \textbf{G, H,} The real part of the density matrices for even tripartite and quadripartite cat states with $\alpha = 0.72$, conditioned on the qubit measured in $\ket{0}$. The matrix elements of the corresponding ideal states are indicated with the transparent boxes. The fidelity of the flying cat states compared to the corresponding ideal states are also labeled.}
\label{cat}
\end{figure*}

\subsection{Multipartite cat states}

Following the above mentioned protocol, we first showcase the preparation of single-party cat states by reflecting a coherent pulse of microwave photons from the cavity when the qubit is prepared at $(\ket{0}+\ket{1})/\sqrt{2}$. We perform quantum state tomography on the reflected itinerant microwave photons conditioned on the qubit state projected to either $\ket{0}$ or $\ket{1}$ with a homodyne setup, considering the added noises introduced by the inefficiency of the detection chain~\cite{Eichler2011,Eichler2012}. In Fig.~\ref{cat}A and B we present the reconstructed Wigner functions for even and odd single-party cat states with $\alpha = 1.07$.
Compared with the corresponding ideal states, we obtain a fidelity of $0.75$ for the even cat state and $0.69$ for the odd cat state. As a characteristic of `non-classicality' of a quantum state, the negative-valued Wigner function reaches about $-0.031\pm0.006$ for the even cat state and $- 0.124\pm0.019$ for the odd cat state, which verifies its non-classical nature. The infidelity of the prepared single-party cat states mainly originates from the photon loss during the reflection process and the qubit decay/dephasing during the cat states preparation.

We further scale the multipartite cat state up to four photonic modes by successively sending coherent photon pulses to the cavity with $\alpha=0.72$ in each pulse. The photon pulses are $500$ ns square pulses and the interval between the adjacent pulses is 300 ns. The full density matrix of the qubit-state-dependent itinerant microwave field is reconstructed with the maximum likelihood quantum state estimation in Fock basis~\cite{Eichler2012,shang2017superfast,besse2020realizing}. Considering the fact that the ideal states are Greenberger-Horne-Zeilinger (GHZ) type entangled coherent states (see Eq.~\ref{q-ph-pi2}), the density matrices are plotted in the cat basis for clarity.
It is worth noting that the population outside this subspace is small enough for an intuitive comparison between the experimental state and the ideal state, as discussed in the Methods.
Specifically, the qubit-state-dependent multipartite cat state has the same parity as is the case for single-party cat, with even (odd) cat state conditioned on the qubit state in $\ket{0}$ ($\ket{1}$).
In Fig.~\ref{cat}C and D we present the experimentally reconstructed density matrices for bipartite cat states, with a fidelity of $0.66$ for even state and $0.62$ for odd state.
Fig.~\ref{cat}G and H show the reconstructed density matrices for even multipartite cat states, with a fidelity of $0.54$ for the tripartite cat state and $0.53$ for the quadripartite cat state, corresponding to a maximally achieved cat size $n(2\alpha)^2$ up to 8.3 photons~\cite{Haroche2008}. 

The dominant error sources for the multipartite cat states are the cavity loss and the qubit decay/dephasing. We note that the fidelity of the multipartite cat state decreases with an increasing number of parties $n$. The cavity loss induced state preparation error is positively related with the cat size, and thus the number of parties. The qubit dephasing induced error is related with the total state preparation time, which is also positively correlated with $n$. These errors can be effectively suppressed by reducing the cavity internal loss and improving the qubit coherence. Taking experimentally achievable improvements, we estimate that it is possible to prepare a multipartite cat state with more than 15 photonic modes and a fidelity better than 0.9 for $\alpha =0.5$, or with 9 photonic modes for $\alpha =2$ with a fidelity of about 0.91, indicating an excellent scalability of the scheme. A detailed discussion about the error model and error budget can be found in the Methods and in Section.~\uppercase\expandafter{\romannumeral4} of the Supplementary Materials.

From the experimental results, we find that odd cat state suffers more error than that for even cat state. A possible explanation is that both the cavity loss induced error and qubit state induced error lead to a mixed state $(\ket{\alpha}\bra{\alpha})^{\otimes n}/2+(\ket{-\alpha}\bra{-\alpha})^{\otimes n}/2$, while an even cat state $\mathscr{N}(\ket{\alpha}^{\otimes n}+\ket{-\alpha}^{\otimes n})$ has a higher fidelity with this mixed state than that for an odd cat state $\mathscr{N}(\ket{\alpha}^{\otimes n}-\ket{-\alpha}^{\otimes n})$

\begin{equation}
\begin{split}
F_0^n(\alpha)&=\frac{1+\exp(-4n|\alpha|^2)+2\exp(-2n|\alpha|^2)}{2+2\exp(-2n|\alpha|^2)},\\
F_1^n(\alpha)&=\frac{1+\exp(-4n|\alpha|^2)-2\exp(-2n|\alpha|^2)}{2-2\exp(-2n|\alpha|^2)}.
\end{split}
\label{cat-mix-fid}
\end{equation}
In Eq.~\ref{cat-mix-fid}, the difference terms scale exponentially with the cat size $n|2\alpha|^2$ predicting a reduced infidelity difference between even and odd cat states with larger sizes.

As discussed before, multipartite states with more superposition and entanglement structures can be synthesized by rotating the qubit to different states before the reflections of coherent state photon pulses. As a proof of principle we prepare $\ket{\alpha}\ket{-\beta} \pm \ket{-\alpha}\ket{\beta}$ type of  cat states by inserting $R_{-y}(\pi)$ rotation between two successive reflection pulses. In the experiment we take $\beta=\alpha$, the resulting density matrix is shown in Fig.~\ref{cat}E and F, with a fidelity of $0.80$ and $0.71$ for the bipartite state conditioned on the qubit in $\ket{0}$ and $\ket{1}$, respectively. From the experimental results, a systematically improved state fidelity can be observed for the cat states with the qubit $\pi$ rotation than that of the corresponding odd or even cat states. This can be explained by the fact that the additional $\pi$ pulse refocuses some of the qubit dephasing during the cat state preparation, leading to an effectively longer qubit dephasing time, and thus smaller errors. Such a refocusing effect can be well produced from the numerical simulation in Section.~\uppercase\expandafter{\romannumeral4} of the Supplementary Materials.
These results demonstrate the great potential of this scheme for the preparation of generalized multipartite cat states more than merely even or odd cat states.

\subsection{Multipartite quantum entanglement}

\begin{figure}[!tbp]
\centering
\includegraphics[width=1\linewidth]{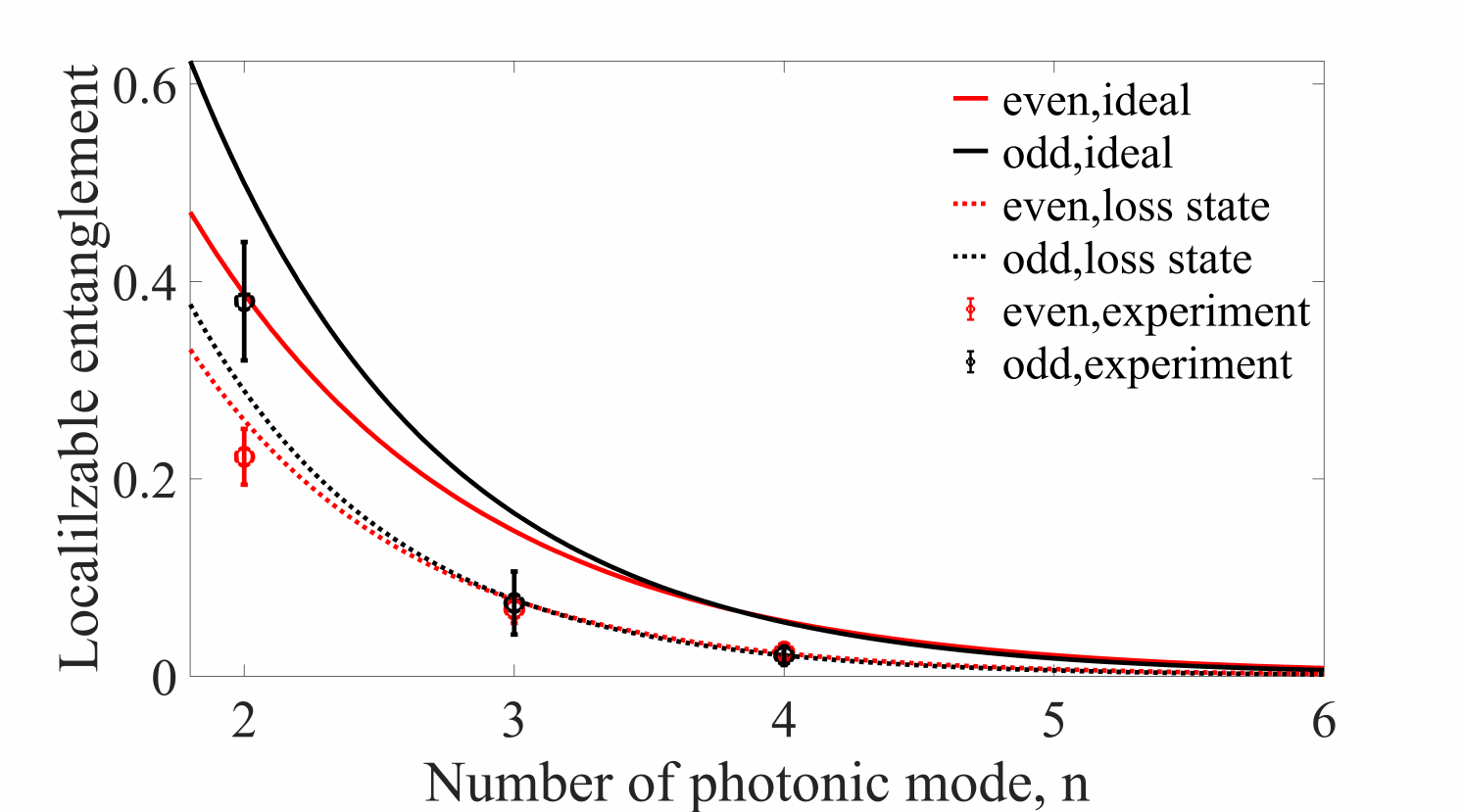}
\caption{\textbf{Localizable entanglement of multipartite cat states.} The deduced localizable entanglement as a function of the number of photonic modes $n$. Localizable entanglement is calculated by projecting the photonic modes apart from the first and the last one to the eigenstates of the operator $x=(a+a^\dagger)/\sqrt{2}$ and calculating the negativities based on the reduced bipartite density matrices. The results for even or odd cat states (corresponding to conditioned qubit state $\ket{0}$ and $\ket{1}$, respectively) are given in red or black, respectively. The scatter plots give the experimental results. The dashed lines are the theoretical results of the corresponding ideal states. The solid lines give the theoretical results considering the possible experimental loss and decoherence. The error bars are extracted from the standard deviation of repeated measurements with a sampling number of $3\times10^7$.}
\label{multipartite}
\end{figure}

In order to verify the existence of quantum entanglement of the prepared states, we calculate localizable entanglement between the first and the last itinerant photonic modes. The full density matrix is reduced by projecting other photonic modes into the eigenstates of the operator $x=(a+a^\dagger)/\sqrt{2}$, leaving a set of bipartite matrices. Note that these projections are purely mathematical operations on the measured density matrix, but not real measurement operations using homodyne detections. In this way we estimate a lower bound on the localizable entanglement~\cite{Cirac2004,besse2020realizing}. An optimized lower bound on the localizable entanglement shall be given by comparing over all possible local measurement strategies, which is beyond the scope of this work. We calculate negativity with the partial transpose of the reduced density matrix, and the localizable entanglement $N$ is quantified by the weighted summation of the negativities of these matrices, which is a metric for the existence of quantum entanglement if $N$ is larger than zero (see more details in Section.~\uppercase\expandafter{\romannumeral3} E of the Supplementary Materials). In Fig.~\ref{multipartite}, the scatter plots show the localizable entanglement of the experimentally prepared multipartite cat states. The statistically positive values clearly demonstrate the existence of quantum entanglement in the multipartite flying photonic states. We note that the calculated localizable entanglement for an ideal multipartite cat state of $n=4$ is close to zero, which does not necessarily mean the absence of entanglement, but is mainly due to the loose lower bound taken in this work. An optimized lower bound can be acquired by choosing a more suitable projection basis~\cite{Cirac2004,Cirac2005}.

\begin{figure}[!tbp]
\centering
\includegraphics[width=1\linewidth]{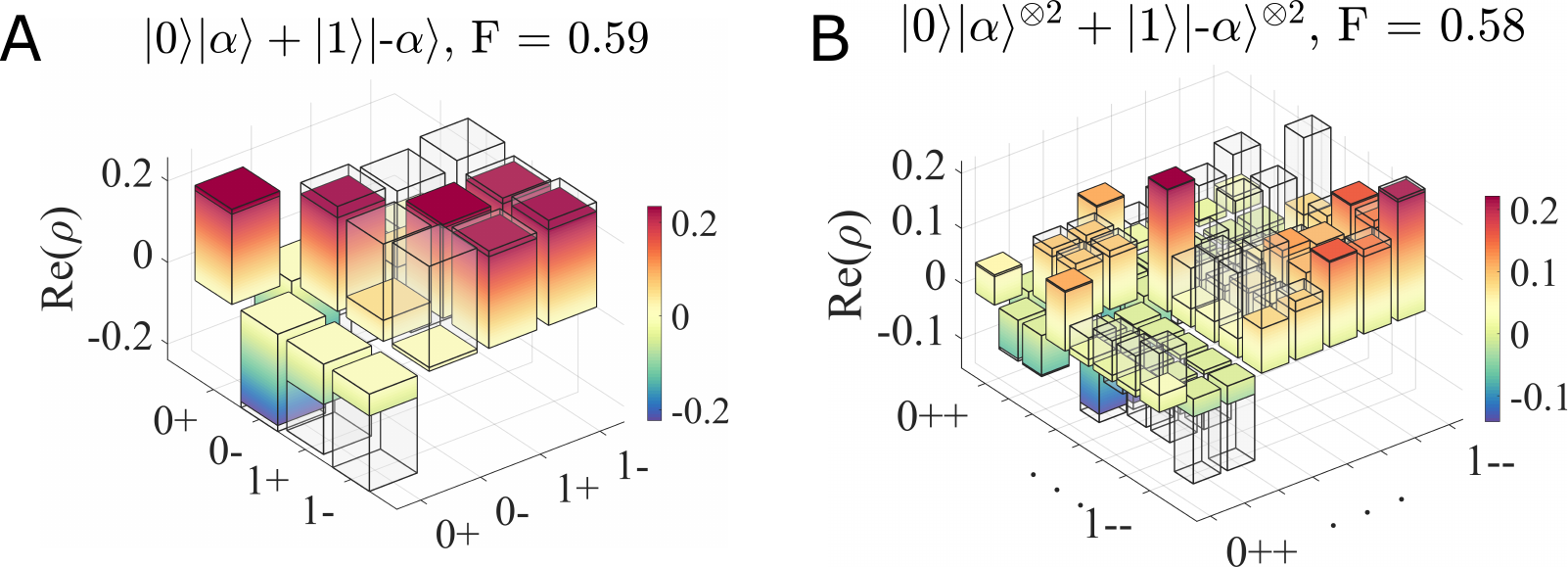}
\caption{\textbf{Quantum entanglement of Schrödinger's cat with a qubit.} \textbf{A,} The real part of the measured joint density matrix for the qubit-photon Schrödinger cat state. \textbf{B,} The real part of the measured joint density matrix for the qubit-photon-photon Schrödinger cat state, which is an entangled state of the qubit and bipartite flying cat state. The matrix elements of the corresponding ideal states are indicted with the transparent boxes. The fidelity of the flying Schrödinger cat states compared to the corresponding ideal states are also labeled. The photonic part of the of the hybrid system is plotted in cat basis for convenience.}
\label{entanglement}
\end{figure}

\subsection{Hybrid quantum entanglement}

The Schrödinger cat state essentially describes hybrid entanglement between discrete-variable states (the superconducting qubit) and multipartite continuous-variable states (itinerant microwave photon fields)~\cite{Furusawa2015,Vlastakis2015,Sun2020}, as shown in Eq.~\ref{qph-entan}. We measure entanglement witness between the superconducting qubit and the itinerant microwave photonic modes based on a joint quantum state tomography~\cite{Eichler2012}. By replacing the $\pi/2$ qubit rotation in Fig.~\ref{illustration}B with $R_{-y}(0),R_{-y}(\pi/2),R_{-x}(\pi/2),R_{-y}(\pi)$, one can get the set of joint moments $\left<a^\dagger a\right>,\left< a^\dagger a\sigma_{x}\right>,\left< a^\dagger a\sigma_{y}\right>,\left< a^\dagger a\sigma_{z}\right>$, respectively. The density matrix of the experimentally prepared qubit-cat system can thus be obtained with a maximum likelihood state estimation~\cite{Eichler2012}, for which the photonic part density matrix is plotted in cat basis for convenience. 
As shown in Fig.~\ref{entanglement}A and B, compared with the corresponding ideal states we measure a fidelity of $0.59$ for the qubit and single-party cat hybrid system, and a fidelity of 0.58 for the qubit and bipartite cat system. The non-vanishing off-diagonal elements in the density matrices demonstrate a coherent superposition of the qubit and the microwave cat state.
In order to witness the existence of quantum entanglement in the hybrid system, we calculate negativity $N$ based on the reconstructed density matrix, as $N_{1}=$ 0.192$\pm0.014$ for the qubit and single-party cat system and $N_{2}=$ 0.135$\pm0.010$ for the qubit and bipartite cat system. The significant positive values unambiguously verify quantum entanglement between the qubit state and the continuous-variable photonic modes.

\section{Conclusion}

In summary, we have presented a highly scalable approach to deterministically generate flying multipartite cat states in microwave regime. By reflecting coherent state photon pulses from a microwave cavity containing a superconducting qubit, we have successfully prepared flying multipartite cat states of microwave photons containing up to four photonic modes. The existence of quantum entanglement in the multipartite cat states and in the qubit-cat hybrid states is verified through a full quantum state tomography of the qubit state and the multipartite photonic states. Multipartite cat states with improved fidelity and larger scale can be obtained by reducing the cavity loss and improving the qubit coherence. In addition, considering that error correction codes~\cite{Leghtas2013,Loock2016,Ofek2016} can be implemented on those cat states, one could envision an extensive application of multipartite cat states in quantum technologies, including demonstration of various quantum metrology and quantum information processing protocols based on cat states~\cite{Munro2002,Jeong2001,Ralph2003,Loock2008,Sangouard2010, Fink2020,Loock2016,Jiang2017prl}.

\bigskip

\section*{Methods}

\textbf{Sample parameters.} The superconducting qubit is made from an aluminum film on a sapphire substrate with standard micro-fabrication techniques. The qubit is placed at the center of a three dimensional aluminum rectangular microwave cavity. The out-coupling rate of the cavity can be precisely tuned by adjusting the length of a one-dimensional transmission line extended into the cavity. The sample is cooled to $T\sim20\,$mK in a dilution refrigerator for measurements.

The energy relaxation time ($T_1$) and coherence time ($T_2$) for qubit excited state $\ket{1}$ is about $20\,\mu s$ and $6\,\mu s$, respectively. The cavity dispersive shift is measured to be $\chi/2\pi = -1.05\,$MHz. The cavity out-coupling rate is tuned to be $\kappa_r/2\pi =2.23\,$MHz, and the internal loss rate is measured as $\kappa_i/2\pi = 0.22\,$MHz.
The coupled system fulfills the optimal condition of $\kappa_{tot}\approx 2|\chi|$. In Fig.~\ref{illustration}C in the main text, one can see that at the bare frequency of the cavity (indicated by the dashed line), the reflection phase difference when the qubit is in $\ket{0}$ and $\ket{1}$ reaches $\pi$, which agrees well with theoretical prediction based on our device parameters (the black curve).
The qubit state readout fidelity in our experiment is determined to be $[P(0|0)+P(1|1)]/2=97.0\%$, where the readout error for the qubit in the ground state (or in the excited state) is measured to be $\epsilon_0=1-P(0|0)=2.69\%$  (or $\epsilon_1=1-P(1|1)=3.27\%)$.

The cavity reflection signal is amplified successively by a Josephson junction parametric amplifier (JPA)~\cite{JPA1,JPA2}, a HEMT amplifier, and two microwave amplifiers at the room temperature before being acquired with a homodyne setup. The JPA is working in a phase preserving mode with a gain of 16 dB around the cavity frequency, yielding an overall circuit detection efficiency $1/(n_{noise}+1)=20\%$ with a noise photon number $n_{noise}=4$.
\bigbreak
\textbf{Quantum state tomography of multipartite cat state.}
The quantum state of the itinerant photonic modes can be reconstructed directly from the measured histograms of the complex amplitudes~\cite{Eichler2012}.
For each of the photonic mode $m$, we measured the complex amplitude $S_m=I_m+iQ_m (m,=1,2,...,n)$, where $n$ represents the number of photonic modes. The measured complex amplitude contains both the signal photons to be characterized and the noise.

First, we need to reconstruct the noise mode $\rho_h$. We prepare the signal photon state as a vacuum state, and measure the complex amplitude $S_{m,vac}$. For each mode $m$, we record a 2-dimensional histogram $D_{m,vac}(\textbf{I},\textbf{Q})$ of the measured complex amplitudes.
The histogram also represents the probability distribution of a positive operator-valued measure (POVM) set $\{\Pi_m^{j,k}\}$, where $\Pi_m^{j,k}=\pi^{-1}\ket{\alpha=I_m^j+iQ_m^k}\bra{\alpha=I_m^j+iQ_m^k}$, here the subscript $m$ indicates the photonic mode, the superscript $j$ and $k$ indicates the $j$th and $k$th bin of the histogram. With the measured probability distribution $D_{m,vac}(\textbf{I},\textbf{Q})$ and the POVM set $\{\Pi_m^{j,k}\}$, we can reconstruct the noise state $\rho_h$ with a conjugate-gradient accelerated projected-gradient (CG-APG) maximum likelihood estimation algorithm~\cite{shang2017superfast}.
The most likely noise state $\rho_h$ given by the measured histogram is the one which maximizes the likelihood function

\begin{equation}
F(\rho_h)=\sum_{j,k,m} D_{m,vac}(\textbf{I},\textbf{Q})\ln(\tr(\rho_h{\Pi_m^{j,k}})).
\end{equation}
In our experiment, the noise mode is reconstructed with a cut-off photon number of $44$, and the reconstructed noise state is very close to a thermal state $\rho_h=\sum_{n=0}^\infty n_{noise}^n/(1+n_{noise})^{n+1}\ket{n}\bra{n}$ with $n_{noise}=4$.

To reconstruct the prepared n-partite cat state, we record the 2n-dimensional histogram $D_p(\textbf{I},\textbf{Q})=D(I_1,Q_1,I_2,Q_2,...,I_n,Q_n)$. The histogram also corresponds to the probability distribution of the POVM set, $\{\Pi_p\}$, where $\Pi_p={\otimes_m}\Pi_{S_m}=\Pi_{S_1=I_1+iQ_1}\Pi_{S_2=I_2+iQ_2}...\Pi_{S_n=I_n+iQ_n}$. In this case $\Pi_{S_m}=\pi^{-1}D_{\alpha=S_m}\rho_h D_{\alpha=S_m}^\dagger$, with $D_{\alpha=S_m}=\exp(\alpha a^\dagger-\alpha^* a)$ as the displacement operator and $\rho_h$ as the reconstructed density matrix of noise mode. Again, the most likely propagating photonic state $\rho_{cat}$ can be reconstructed with the CG-APG maximum likelihood estimation algorithm by maximizing the likelihood function

\begin{equation}
F(\rho_{cat})=\sum_{p}D_p(\textbf{I},\textbf{Q})\ln(\tr(\rho_{cat}{\Pi_p})).
\end{equation}

In the experiment we collect the complex amplitudes of the itinerant photonic modes with a sampling number of $3\times10^7$. As for the state reconstruction, we use a cut-off photon number of 8 for single-party, bipartite and tripartite cat states, and a cut-off photon number of 5 for quadripartite cat cases. Note that we use $\alpha =0.72$ for the generation of multipartite cat states, a cut-off photon number of 5 is much larger than the average photon number of each modes $|\alpha|^2 = 0.52$, which indicates that photon number truncation is sufficient enough.

\bigbreak
\textbf{Joint tomography of Schrödinger cat states and a qubit.}
A joint quantum state tomography is applied to the entangled system in order to characterize the quantum entanglement between the qubit and the flying cat~\cite{Eichler2012}.
The basic idea is to measure the photon field when projecting the qubit state in different bases. To this aim, we perform the measurement in four different qubit bases by using different unitary qubit rotations $R_{\phi_{b}}(\theta_{b})$, instead of the $\pi/2$ pulse in Fig.~\ref{illustration}B, as $R_{-y}(0),R_{-y}(\pi/2),R_{-x}(\pi/2),R_{-y}(\pi)$, accordingly the qubit is measured by Pauli matrices $\{\sigma_I,\sigma_x,\sigma_y,\sigma_z\}$.

 We can calculate the moments $\langle(S^\dagger)^mS^n\rangle |\sigma_i^0$ and $\langle(S^\dagger)^mS^n\rangle|\sigma_i^1$ for the experimentally measured complex amplitudes $S=I+iQ$, where $\sigma_i^{0(1)}$ means that the specific moment is calculated with the complex amplitude $S$ when the qubit is measured by Pauli matrix $\sigma_i(i=I,x,y,z)$ with a measurement result of $\ket{0}$ (or $\ket{1})$~\cite{Eichler2012}.

 The measured $S$ contains both the reflected photonic modes $a$ and the noise $h$. We can get photon state moments $\langle(a^\dagger)^ma^n\rangle |\sigma_i^0$ and $\langle(a^\dagger)^ma^n\rangle|\sigma_i^1$ by removing the influence of the noise~\cite{Eichler2012}. The final qubit-photon joint moments $\langle(a^\dagger)^ma^n\sigma_i \rangle$ can thus be calculated by $\langle(a^\dagger)^ma^n\sigma_i \rangle=P_0\lambda_0 \langle(a^\dagger)^ma^n\rangle |\sigma_i^0+P_1\lambda_1\langle(a^\dagger)^ma^n\rangle |\sigma_i^1$, where $P_{0(1)}$ is the probability of the qubit state measured in $\ket{0}$ (or $\ket{1})$ under $\sigma_i$, $\lambda_{0(1)}$ is the eigenvalue of the measurement result for $\sigma_i$. For $\sigma_I$, $\lambda_0=\lambda_1=1$, while for other Pauli matrices, $\lambda_0=1$ and $\lambda_1=-1$.
Finally, by using the maximum-likelihood method with the log-likelihood function
\begin{equation}
L_{log}=-\sum_{n,m,i}\frac{1}{\delta_{m,n,i}^2}|\langle(a^\dagger)^ma^n\sigma_i\rangle-\Tr[\rho_{q-cat}(a^\dagger)^ma^n\sigma_i]|^2,
\end{equation}
the experimental state $\rho_{q-cat}$ can thus be reconstructed, where $\delta_{m,n,i}$ is the standard deviation of $\langle(a^\dagger)^ma^n\sigma_i \rangle$ obtained from repetitive measurements.
For the entangled state between the qubit and two photons, we use the same method but using the moments of two photonic modes $\langle(S_1^\dagger)^mS_1^n(S_2^\dagger)^iS_2^j\rangle$.
In the experiment the complex amplitudes of the photonic modes for each of the four qubit bases are sampled $3\times10^7$ times. We calculate the moments up to $m+n = 6$ (or $m+n+i+j = 6$) and use a cut-off photon number of 9 (7) for the reconstruction of the qubit and single-party cat (bipartite cat) entangled states.

\bigbreak

\textbf{Cat basis representation.}
As can be seen from the discussion on state tomography, the state space would be very large for the multipartite cat state in Fock basis. For example, the reconstruction of quadripartite cat state with a cut-off photon number of 5 yields a state space dimension of 1296, which is inconvenient for visualization. In Fig. S5 in the Supplementary Materials, we plot the density matrix of even cat states in Fock basis, in which the non-zero matrix elements are already very blurred and hard to be compared. In order to have an intuitive and clear comparison between the experimental state and the ideal state, especially considering the fact that the multipartite cat state describes a GHZ type entanglement of coherent states, we plot the density matrix obtained in Fock basis to a coherent state based cat basis. The cat basis can be given as $\{\ket{\pm_1\pm_2\cdots\pm_n}=(\ket{\alpha_1}\pm\ket{-\alpha_1})\otimes(\ket{\alpha_2}\pm\ket{-\alpha_2})\otimes\cdots\otimes(\ket{\alpha_n}\pm\ket{-\alpha_n})\}$, where $\alpha_i$ ($i=1,2...,n$) represents the $i$th coherent mode, and $n$ is the number of photonic modes.

It is worth noting that the cat basis is orthogonal but not complete, which means some population may leaks out of the subspace resulting in $\Tr{\rho_{cat}}\le1$. For single-party even (odd) cat state, we estimate a trace in the subspace is about $94\%$ ($92.5\%$), and for quadripartite even (odd) cat state we estimate a trace in the subspace is about $80.6\%$ ($73.6\%$). Considering the fact that we only use the cat basis for visualization, such a completeness of the cat basis representation is already enough to make an intuitive comparison between the experimental state and the ideal state.

\bigbreak
\textbf{Error budget.}
We have developed an error model for the cat state preparation process (see Section.~\uppercase\expandafter{\romannumeral4} A of Supplementary Materials for details), based on which we can make an error budget for the experimental cat states, as shown in the Table S2 in the Supplementary Materials, which includes the infidelity induced by cavity loss during the reflection process, the qubit decay/dephasing and qubit state measurement error.

From the Table S2 in the Supplementary Materials we could find that the cavity-loss-induced error contributes a lot in the total infidelity of the prepared states, especially for multipartite cat states or cat states with a large size.
A direct way to improve the fidelity of the generated state is to reduce the cavity loss, which is crucial for the generation of multipartite cat states or cat states with a large size and a high fidelity.
The internal loss rate of the cavity used in the experiment is $\kappa_i = 0.22\,$MHz, which may be induced by the surface defects of our cavity, or by the coupled qubit, including substrate loss, and mechanical instability. If $\kappa_i$ can be reduced by an order of magnitude to about $20\,$kHz, the cavity loss induced error can be reduced by an order of magnitude, from about $0.2$ for single-party even cat and $0.37$ for quadripartite even cat to about $0.01$ for single-party case and $0.02$ quadripartite case (see Section.~\uppercase\expandafter{\romannumeral4} B of Supplementary Materials for details). Note that reducing $\kappa_i$ to around $20\,$kHz is doable considering that the state of the art cavity loss can be well controlled to below $1\,$kHz by chemical etching~\cite{Reagor2016}.

In the Table S2 in the Supplementary Materials, one could find that the qubit decay/dephasing induced infidelity increases with increasing number of parties, which is easy to understand considering the fact that the state preparation time gets longer with increasing number of parties, and thus the qubit suffers more decay and dephasing, leading to a cat state with more error.
In our experiment the qubit state induced error is dominated by qubit dephasing, since for our sample $T_2$ is closer to the experimental sequence duration and much shorter than $T_1$.


\bigskip

\textbf{Acknowledgements:} The authors thank Wei Zhang for providing us with the cat cartoon. This work was supported by the Frontier Science Center for Quantum Information of the Ministry of Education of China through the Tsinghua University Initiative Scientific Research Program, the National Natural Science Foundation of China under Grant No.11874235 and No.11925404, the National key Research and Development Program of China (2017YFA0304303, 2020YFA0309500), the Key-Area Research and Development Program of Guangdong Province (No.2020B0303030001) and a grant (No.2019GQG1024) from the Institute for Guo Qiang, Tsinghua University. Y.K.W. acknowledges support from Shuimu Tsinghua Scholar Program and the International Postdoctoral Exchange Fellowship Program.

\textbf{Author Contributions:} L.M.D. and H.Y.Z. proposed and supervised the experiment. Z.L.W., Z.H.B., W.Z.C. and L.Y.S. prepared the sample. Z.L.W., Z.H.B., Y.L., W.T.W., Y.W.M, T.Q.C., X.Y.H., J.H.W., Y.P.S. and H.Y.Z. collected and analyzed the data. Z.L.W., Z.H.B. and Y.K.W. carried out the theoretical analysis. Z.L.W., Z.H.B., Y.K.W., H.Y.Z. and L.M.D. wrote the manuscript.

\textbf{Competing interests:} The authors declare that there are no competing interests.

\textbf{Data Availability:} The data that support the findings of this study are available from the authors upon request.

\end{document}